\documentclass{iau} 
\usepackage{graphicx}
\usepackage[compact]{titlesec}
\usepackage{setspace}

\title[Galaxy evolution studies in clusters] 
{Galaxy evolution studies in clusters: the case of Cl0024+1652 cluster galaxies at $z$\,$\sim$\,0.4}

\author[Z. Beyoro-Amado et al.]   
{Zeleke Beyoro-Amado$^{1,\,2,\,3}$,\,\,
Mirjana Povi\'c$^{1,\,4}$,  \\
Miguel S\'anchez-Portal$^{5}$,\,\,  
Solomon Belay Tessema$^{1}$, \\Tilahun Getachew-Woreta$^{1,\,2,\,6}$ \and 
the GLACE team}

\affiliation{$^{1}$Ethiopian Space Science and Technology Institute (ESSTI), Entoto Observatory and Research Centre (EORC),
Astronomy and Astrophysics Research and Development Division, P.O.Box 33679, Addis Ababa, Ethiopia\\
$^{2}$Addis Ababa University (AAU), P.O.Box 1176, Addis Ababa, Ethiopia\\
$^{3}$Kotebe Metropolitan University, College of Natural and Computational Sciences, Department of Physics, 
P.O.Box 31248, Addis Ababa, Ethiopia\\
$^{4}$Instituto de Astrof\'isica de Andaluc\'ia (IAA-CSIC), Glorieta de la Astronom\'ia s/n, 18008, Granada, Spain\\
$^{5}$Instituto de Radioastronom\'ia Milim\'etrica, Av. Divina Pastora 7, N\'ucleo Central, E-18012 Granada, Spain\\
$^{6}$Bule Hora University, P.O.Box 144, Bule Hora, Ethiopia\\
}

\pubyear{2019}
\volume{356}  
\jname{Nuclear Activity in Galaxies Across Cosmic Time}
\editors{M. Povi\'c, P. Marziani, J. Masegosa, H. Netzer, \\P. Shastri, S.H. Negu, $\&$ S.B. Tessema, eds.}
\begin{document}

\maketitle

\begin{abstract}
Studying the transformation of cluster galaxies contributes a lot to have a clear picture of evolution of the universe. Towards that we are studying different properties (morphology, star formation, AGN contribution and metallicity) of galaxies in clusters up to $z$\,$\sim$\,1.0 taking three different clusters: ZwCl0024\,+\,1652 at $z$\,$\sim$\,0.4, RXJ1257\,+\,4738 at $z$\,$\sim$\,0.9 and Virgo at $z$\,$\sim$\,0.0038. For ZwCl0024\,+\,1652 and RXJ1257\,+\,4738 clusters we used tunable filters data from GLACE survey taken with GTC 10.4\,m telescope and other public data, while for Virgo we used public data. We did the morphological classification of 180 galaxies in ZwCl0024\,+\,1652 using galSVM, where 54\,\% and 46\,\% of galaxies were classified as early-type (ET) and late-type (LT) respectively. We did a comparison between the three clusters within the clustercentric distance of 1\,Mpc and found that ET proportion (decreasing with redshift) dominates over the LT (increasing with redshift) throughout. We finalized the data reduction for ZwCl0024\,+\,1652 cluster and identified 46 [OIII] and 73 H$\beta$ emission lines. For this cluster we have classified 22 emission line galaxies (ELGs) using BPT-NII diagnostic diagram resulting with 14 composite, 1 AGN and 7 star forming (SF) galaxies. We are using these results, together with the public data, for further analysis of the variations of properties in relation to redshift within $z$\,$<$\,1.0.      
\end{abstract}
\section{Introduction}
The way galaxies form and evolve in both field and clusters still remains one of the open questions in modern cosmology. Understanding how galaxies transform inside the clusters presents one of the main steps in disentangling the picture of galaxy formation and evolution, and the formation of the universe at large. In this regard significant differences have been observed between the field and cluster galaxies (e.g., \cite[Koopman \& Kenny 1998]{KoopKen1998}).

Research results show that significant evolution in properties of galaxies within clusters are observed as a function of redshift.  
According to previous studies, the cores of low redshift clusters are dominated by red ET galaxies while the blue LT galaxy population dominates the higher redshift ($z$) clusters (e.g., \cite[Butcher \& Oemler 1984]{ButOem1984}, \cite[Beyoro-Amado et al. 2019]{Amado2019}). Concerning the star formation activity, an increase of the obscured SF in mid-infrared (MIR) and far-infrared (FIR) surveys of distant clusters has been indicated (e.g., \cite[Haines et al. 2009]{Hai09}). In addition to these, higher population of AGN have been observed at higher redshifts 
(e.g., \cite{Bufanda2017}). 

\cite{KoB01} and \cite{Oh2018} described that properties of galaxies in clusters also transform significantly with environment. Different physical processes were suggested for affecting the galaxy evolution in clusters. According to \cite{Treu2003} tidal halo stripping, tidal triggering star formation, and ram pressure stripping affect the galaxy evolution in the cluster core ranging to about 2\,Mpc from the cluster center while the effects of starvation, harassment and merging remain significant up to a clustercentric distance of about 5\,Mpc. 

Morphological classification of galaxies could be performed either by traditional 
(visual inspection; e.g., \cite{Lintott2011}) or modern techniques. The modern techniques can be either parametric, performed by fitting some parameters assuming a predefined parametric model (e.g., \cite{Tarsitano2018}), or non parametric, where no specific analytic model is assumed while performed by measuring a set of well-chosen observables (e.g., \cite[Povi\'c et al. 2015]{Povic2015}, \cite[Huertas-Company et al. 2015]{HC2015}, \cite[Pintos-Castro et al. 2016]{PC2016}, \cite[Beyoro-Amado et al. 2019]{Amado2019}). Previous studies show that local density has effects on galaxy morphology (e.g., \cite{Nantais2013}); as well as SF activities (e.g., \cite{Woo2013}). Generally speaking, galaxy properties like morphology, nuclear activity, SF activity, metallicities, and color vary in relation to redshift and environment (e.g., \cite{Lagana2018}).

Several controversial results exist in relation to AGN activities in clusters. For instance there are studies describing that AGN activities do not depend on local densities (e.g., \cite{Miller2003}) in one hand while there are others indicating that high density regions avoid luminous AGNs (e.g., \cite{Powell2018}).
AGN fraction versus environment, metallicities, population of ELGs in relation to radial distance, and morphological transformations with radial distance and redshift for galaxy clusters still remain some open issues. This motivated us to carry out a PhD research (for the first author here) on galaxy evolution and transformation in clusters by studying different properties of three galaxy clusters. 

Our work is generally aimed at studying the properties of galaxies in clusters up to $z$\,$\sim$\,1.0 using three galaxy clusters: ZwCl0024\,+\,1652 at $z$\,$\sim$\,0.4, RXJ1257\,+\,4738 at $z$\,$\sim$\,0.9 and Virgo at $z$\,$\sim$\,0.0038. Hence performing the detailed morphological study, and analysing the AGN versus SF contributions plus the metallicity variations; could give us important clues about galaxy evolution and transformations with cosmic time.

\section{Data}
For ZwCl0024\,+\,1652 cluster, we used the tunable filter (TF) raw observational data from GaLAxy Cluster Evolution (GLACE) survey with targets to [OIII], H$\beta$, H$\alpha$ and [NII] emission lines (\cite{Sanchez2015}). The GLACE data were taken with Gran Telescopio Canarias (GTC) 10.4m telescope located at La Palma. For this cluster, we also used HST/ACS data, WFP2 master catalogue (\cite{Moran2005}), and visual morphology catalogue (\cite{Moran2007}). For the cluster at $z$\,$\sim$\,0.9, we used the GLACE TF data targeting H$\beta$ and [OII] lines, the FIR data (\cite{PC2013}) and the catalogue of morphological properties (\cite[Pintos-Castro et al. 2016]{PC2016}). We used public data in case of Virgo cluster: the Virgo cluster catalogue (VCC, \cite{Binggeli1985}), HST/ACS images (\cite{Peng2008}), Herschel reference survey data (HRS, \cite{Hughes2013}), and extended Virgo cluster catalogue data (EVCC, \cite{Kim2014}).      

\section{Results of morphological study}
We have performed the morphological study for all the three clusters and did a comparison among them.
For the morphological classification of ZwCl0024+1652, we used a non parametric public code called \underline{gal}axy \underline{S}upport \underline{V}ector \underline{M}achine (galSVM) operating in IDL environment. The code simultaneously measures six morphological parameters: Abraham concentration (CABR), moment of light (M20), Bershady-Conselice concentration (CCON), Gini coefficient (GINI), asymmetry (ASYM) and smoothness. Using these parameters together with ellipticity from SExtractor, galSVM gives a probability ($P_{ET}$) for the galaxy to be an ET where the probability for being LT is $1-P_{ET}$. We measured morphological parameters for 231 member galaxies within the clustercentric distance of 1\,Mpc, where 180 galaxies have been classified with 97 ($\sim$\,54\%) and 83 ($\sim\,$46\%) galaxies have been classified as ET and LT respectively. The classification is in $\sim\,$80\% agreement with previous works (\cite{Moran2007}) and 121 new galaxies have been classified. With this work we obtained one of the most complete morphological catalogue of ZwCl0024\,+\,1652 cluster within the clustercentric radius of 1\,Mpc (\cite[Beyoro-Amado et al. 2019]{Amado2019}).         

To track the trends of evolution of cluster galaxies, we compared the results for the three clusters. The morphological class comparison is presented in Table \ref{tab1}.    
\begin{table}[h]
  \begin{center}
  \caption{Variation of morphological classifications of cluster galaxies with redshift}
  \label{tab1}
 {\scriptsize
  \begin{tabular}{|l|l|c|c|}\hline 
{\bf Cluster} & {\bf Redshift ($z$)} & {\bf ET} & {\bf LT} \\ \hline
ZwCl0024\,+\,1652 & $\sim$\,0.4 & 97 ($\sim$\,54\%) & 83 ($\sim$\,46\%) \\ \hline
RXJ1257\,+\,4738 & $\sim$\,0.9 & 31 ($\sim$\,89\%) & 4 ($\sim$\,11\%) \\ \hline
Virgo & $\sim$\,0.0038 & 273 ($\sim$\,68\%) & 131 ($\sim$\,32\%) \\ \hline
  \end{tabular}
  }
 \end{center}
 \end{table}
 From the table we can see that in a redshift range to $z$\,$\sim$\,1 within a clustercentric distance of 1\,Mpc, the proportion of ET galaxies dominate over the LT galaxies. Here because of the small number of sources especially for RXJ1257\,+\,4738 cluster, we couldn't generalize, but from results of ZwCl0024\,+\,1652 and Virgo clusters, the usual evolution of morphologies of cluster galaxies holds in such a way that the proportion of LT increases while ET proportion decreases as redshift increases. 
 \begin{figure}[h]
\begin{center}
\includegraphics[width=5.0in, height=1.8in]{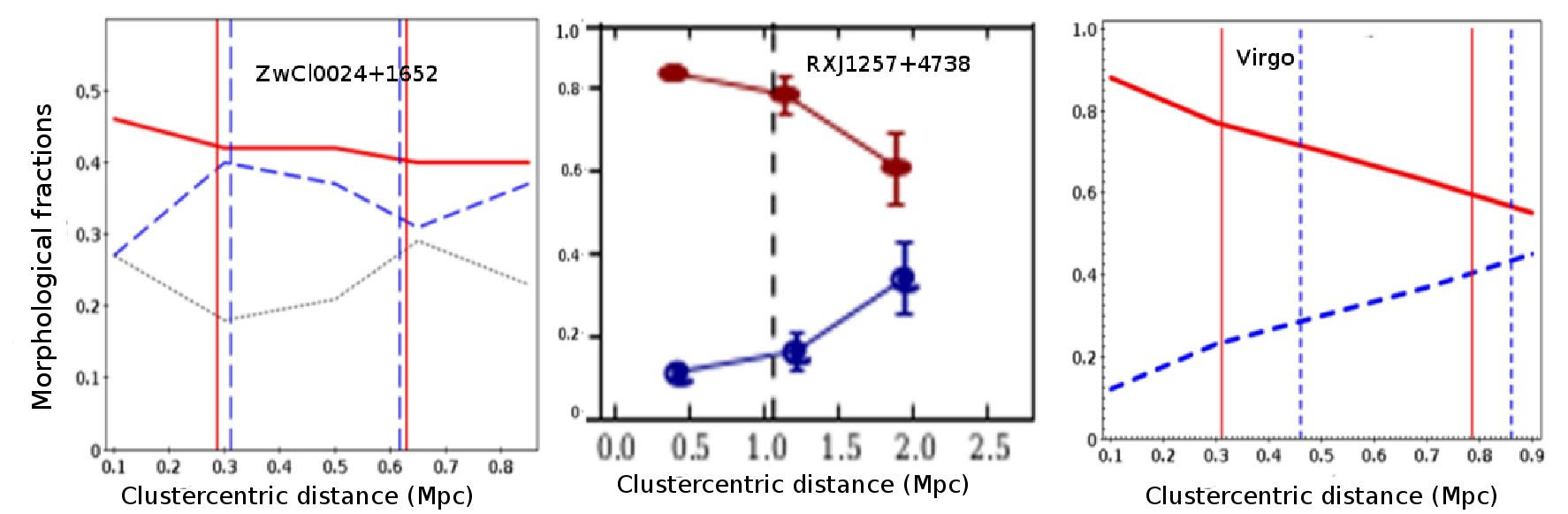} 
\caption{Variations of morphological fractions with clustercentric distance for the three clusters. The red lines stand for ET while the blue lines represent the LT galaxies. The middle plot is recovered from \cite[Pintos-Castro et al. (2016)]{PC2016}.}
   \label{fig1}
\end{center}
\end{figure}
We also present the comparison of the results in morphological fractions for the three clusters at different redshifts as a function of clustercentric distance in Fig.\,\ref{fig1}. The plots show that for clusters at $z$\,$<$\,1, the fractions of ET galaxies decrease with an increase in LT galaxies as a function of clustercentric distance out to 1\,Mpc. It looks that at lower redshifts the morphological fractions change faster (decrease for ET, increase for LT) than in the case of higher redshifts. 
\section{The GLACE TF data reduction for ZwCl0024\,+\,1652 and RXJ1257\,+\,4738 clusters}
\begin{figure}[h]
\begin{center}
\includegraphics[width=4.5in, height=2.0in]{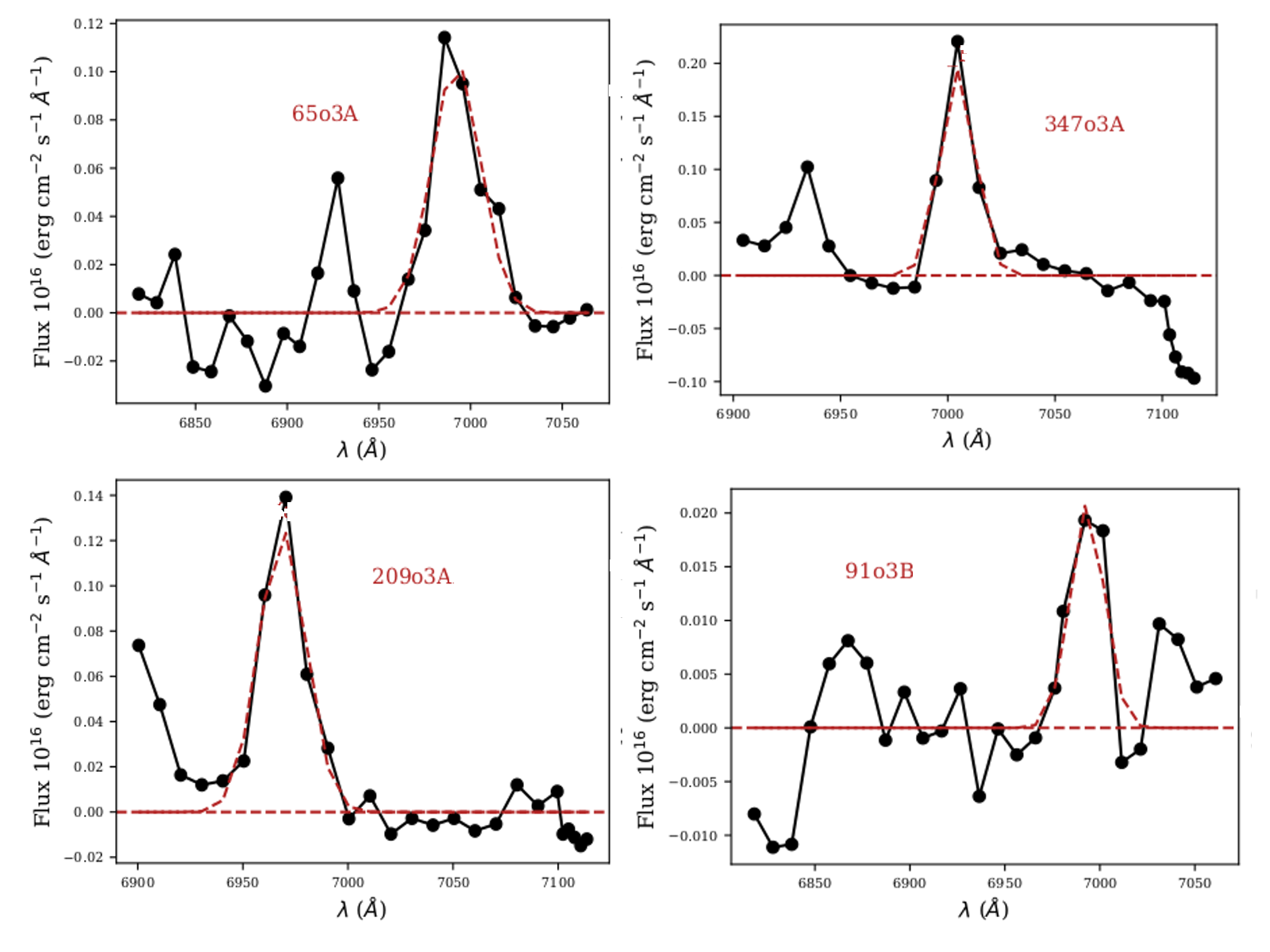} 
\caption{Example of pseudospectra indicating some of the [OIII] emission lines idendified. The red label in each plot indicates the unique id of the galaxy as in our results.}
   \label{fig2}
\end{center}
\end{figure}

We had a TF raw data of ZwCl0024\,+\,1652 and RXJ1257\,+\,4738 clusters observed under GLACE project targeting [OIII] and H$\beta$ lines. For data reduction, the basic standard steps have been implemented and we used different scripts (in IDL, IRAF and python) with TFREd package suited for TF data. We used GAIA astrometry for mapping the cluster before doing the wavelength and flux calibration. Using H$\alpha$ results (\cite{Sanchez2015}) of Zwcl0024\,+\,1652 cluster,  we matched our results, measured the fluxes and plotted the pseudospectra of all sources. Pseudospectra of some of the [OIII] lines are presented in Fig.\,\ref{fig2} as an example. 
By relying on better signal to noise ratio and visual inspection of the pseudospectra, a total of 46 [OIII] and 73 H$\beta$ emission lines have been identified (both increasing with clustercentric distance). 
\begin{figure}[h]
\begin{center}
\includegraphics[width=4.5in, height=2.0in]{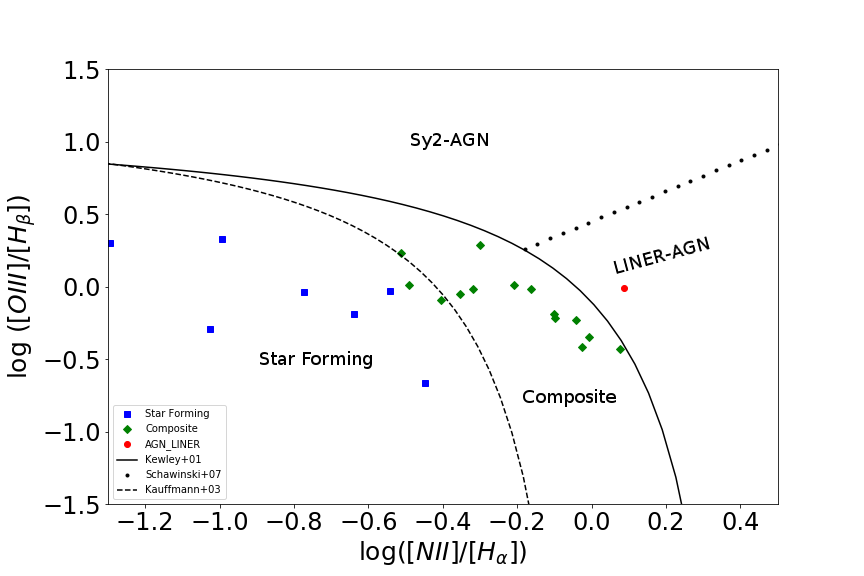} 
\caption{Our results of classifying the ELGs using the BPT-NII diagram}
   \label{fig3}
\end{center}
\end{figure}
Combining ours with the published results (\cite{Sanchez2015}) we have identified 22 galaxies with all four emission lines; ([OIII], H$\beta$, H$\alpha$ and [NII]). For these member galaxies we plotted the BPT-NII diagnostic diagram (\cite{Kew2006}) to classify them into SF (7), composite (14), and LINER-AGN (1), as shown in Fig.\,\ref{fig3}.
\section{Conclusions and the work on progress}
From morphological analysis, we suggested that cores of galaxy clusters at $z$\,$<$\,1.0 are dominated by ET galaxies whose fraction decreases as a function of clustercentric distance up to 1\,Mpc. The LT proportions increase with redshift and separately with the clustercentric distance. We also suggested that ELGs do not populate cluster cores and increase in numbers outwards with clustercentric distance. The majority of ELGs in ZwCl0024\,+\,1652 cluster are SF with small number of AGNs comparing the two classes; while more than half of the ELGs are composite. 
Currently, we are working on the SF/AGN, SFR and metallicity analysis of ZwCl0024\,+\,1652 cluster. Then having the results, comparison with RXJ1257\,+\,4738 and Virgo will be performed to draw final conclusions on the evolution and transformation of galaxies in clusters up to $z$\,$\sim$\,1.0. 

\section*{Acknowledgements}
We thank the Ethiopian Space Science and Technology Institute (ESSTI) under the Ethiopian Ministry of Innovation and Technology (MoIT) for all 
the financial support. ZBA especially acknowledges Instituto de Radioastronom\'ia Milim\'etrica (IRAM) for the financial and logistics supports, and Instituto de Astrof\'isica de Andaluc\'ia (IAA-CSIC) for providing the working space during the research visits for part of this work.


\begin{thebibliography}{}

\bibitem[Beyoro-Amado et al. (2019)]{Amado2019} Beyoro-Amado, Z., Povi\'c, M., S\'anchez-Portal, M., et al., 2019, \textit{MNRAS}, 485, 1528
\bibitem[Binggeli et al. 1985]{Binggeli1985} Binggeli, B., et al., 1985, \textit{AJ}, 90, 1681 
\bibitem[Bufanda et al. 2017]{Bufanda2017} Bufanda, E., Hollowood, D., et al., 2017, \textit{MNRAS}, 465, 2531
\bibitem[Butcher \& Oemler (1984)] {ButOem1984} Butcher, H. \& Oemler, A., 1984, \textit{ApJ}, 285, 426
\bibitem[Haines et al. 2009]{Hai09} Haines, C. P., et al. 2009, \textit{ApJ}, 704, 126
\bibitem[Huertas-Company et al. (2015)]{HC2015} Huertas-Company, M., et al., 2015, \textit{ApJS}, 221:8
\bibitem[Hughes et al. 2013]{Hughes2013} Hughes, T. M., et al. 2013, \textit{A$\&$A}, 550, A115
\bibitem[Kewley et al. 2006]{Kew2006} Kewley, L. J., et al. 2006, \textit{MNRAS}, 372, 961
\bibitem[Kim et al. 2014]{Kim2014} Kim, S., et al. 2014, \textit{ApJS}, 215:22  
\bibitem[Kodama \& Bower (2001)]{KoB01} Kodama, T., \& Bower, R., 2001, \textit{MNRAS}, 321, 18
\bibitem[Koopman \& Kenny (1998)]{KoopKen1998} Koopmann, R. A., \& Kenney, J. D. P. 1998, \textit{ApJ}, 497, L75
\bibitem[Lagan\'a \& Ulmer 2018]{Lagana2018} Lagan\'a, T. F., \& Ulmer, M. P., 2018, \textit{MNRAS}, 475, 523
\bibitem[Lintott et al. 2011]{Lintott2011} Lintott, C. J., et al. 2011, \textit{MNRAS}, 410, 166
\bibitem[Martini et al. 2013]{Martini2013} Martini, P., Miller, E. D., Brodwin, M., et al. 2013, \textit{ApJ}, 768, 1
\bibitem[Miller et al. 2003]{Miller2003} Miller, C. J., et al., 2003, \textit{ApJ}, 597, 142
\bibitem[Moran et al. 2005]{Moran2005} Moran, S. M., Ellis, R. S., Teu, T., et al. 2005, \textit{ApJ}, 634, 977
\bibitem[Moran et al. 2007]{Moran2007} Moran, S. M., Ellis, R. S., Teu, T., et al. 2007, \textit{ApJ}, 671, 1503
\bibitem[Nantais et al. 2013]{Nantais2013} Nantais, J. B., Flores, H., Demarco, R., et al. 2013, \textit{A$\&$A}, 555, A5
\bibitem[Oh et al. (2018)]{Oh2018} Oh S., Kim K., et al. 2018, \textit{AJSS}, 237, 14
\bibitem[Peng et al. 2008]{Peng2008} Peng, E. W., et al., 2008, \textit{ApJ}, 681, 197
\bibitem[Pintos-Castro et al. (2016)]{PC2016} Pintos-Castro, I., Povi\'c, M., et al. 2016, \textit{A$\&$A}, 592, A108
\bibitem[Pintos-Castro et al. 2013]{PC2013} Pintos-Castro, I., S\'anchez-Portal M., et al. 2013, \textit{A$\&$A}, 558, A100
\bibitem[Povi\'c et al. (2015)]{Povic2015} Povi\'c M., M\'arquez I., et al. 2015, \textit{MNRAS}, 453, 1644
\bibitem[Powell et al. 2018]{Powell2018} Powell, M.~C., et al. 2018, \textit{ApJ}, 858, 110
\bibitem[S\'anchez-Portal et al. 2015]{Sanchez2015} S{\'a}nchez-Portal M., Pintos-Castro I., et al. 2015, \textit{A$\&$A}, 578, A30
\bibitem[Tarsitano et al. 2018]{Tarsitano2018} Tarsitano, F., et al. 2018, \textit{MNRAS}, 481, 2018
\bibitem[Treu et al. (2003)]{Treu2003} Treu, T., Ellis, R. S., Kneib, J., et al. 2003, \textit{APJ}, 591, 53
\bibitem[Woo et al. 2013]{Woo2013} Woo J., et al., 2013, \textit{MNRAS}, 428, 3306

\end{thebibliography}
\end{document}